\documentclass[checkin,pre,showpacs]{revtex4}
\begin{document}        %
\draft
\title{Possible Pressure Waves of Supersolid Helium at High Pressure} 
\author{Chu, Kwang-Hua W.} 
\affiliation{Department of Physics, Xinjiang University, 14, Road
Shengli, Urumqi 830046, PR China }
\begin{abstract}
We make comments on Kim and Chan's [{\it Phys. Rev. Lett.} 97,
 115302 (2006)] letter. Based on their pressure-dependent measurements
 (by a torsional oscillator technique),
we propose that the {\it supersolid} fraction ($\rho_s/\rho$) might
be relevant to an sound absorption or attenuation (fluctuations of
pressure waves) in microscopic domain since there is no apparent
change in $T_c$ with pressure.
\end{abstract}
\maketitle
\bibliographystyle{plain}

Quite recently Kim and Chan have measured the pressure dependence of
the supersolid fraction by a torsional oscillator technique.
Superflow is found from 25.6 bar up to 136.9 bar [1]. Kim and Chan
 used oscillation speed ($v_{max}$ : maximum oscillation speed
of the annulus) of 5$\mu$m/s or less to study the supersolid [2-10]
response of 9 additional solid samples at 25.6, 41.8, 48.7, 56.9,
60.1, 70.6, 87.1, 99.0, and 104.0 bar (they found the NCRIF
(nonclassical rotational inertia (NCRI) fraction) is independent of
$v_{max}$, provided $v_{max}$ does not exceed 10 $\mu$m/s. Once
exceeded, NCRIF decreases with $v_{max}$. They thus interpreted this
as a critical velocity effect [3]. NCRIF measured with $v_{max}$
smaller than 10 $\mu$m/s, being independent of oscillation speed,
represents the supersolid fraction, $\rho_s/\rho$). The uncertainty
in pressure determination is less than 0.5 bar. The low temperature
supersolid fractions, $\rho_{so}/\rho$, of all fourteen samples are
plotted in Fig. 4 of [1] as a function of pressure. The
non-monotonic dependence of the supersolid fraction on pressure
indicates that, as commented in [1], the origin of supersolidity is
more subtle than just the simple Bose condensation of zero point
vacancies. The fact that Kim and Chan found a supersolid fraction of
up to 1.5\% is also difficult to reconcile with the simple vacancy
condensation model. A number of experiments [11] give indirect
evidence that zero point vacancies, if present below 0.2K, would be
much smaller than 1\% of the lattice sites.\newline
We know that solid helium at an elevated pressure is expected to be
less quantum mechanical than that at a lower pressure [12]. X-ray
diffraction studies measuring the zero-point energy induced motions
of the $^4$He atoms from their lattice sites appear to confirm this
expectation [13]. The declining supersolid fraction with pressure
beyond 55 bar is also consistent with this expectation. However, Kim
and Chan do not understand why there is no apparent change in T$_c$
with pressure. This could be one evidence that there are pressure
oscillations (fluctuations of waves although being rather small)
during the imposing process of high pressures in [1].
\newline
The present author also noticed that broad (dissipation) maxima
centering near where NCRIF is changing most rapidly were found in
[1] (cf. Fig. 2 therein). These broad maxima in dissipation are more
pronounced in low pressure solid samples and in data taken at low
$v_{max}$. The dissipation maximum fades with increasing pressure
and it is barely discernible in samples with pressure exceeding 108
bar. This behavior, as the present author compared it with those in
[14], looks like an acoustic  absorption or attenuation
(fluctuations of high-frequency pressure waves) in a rarefied
environment.
Meanwhile, Rittner and Reppy just reported  supersolid decoupling in
a solid sample made by the same blocked capillary method [10].
However, upon annealing the sample by cooling it much more slowly
from about 1.5K than when it was first grown, the supersolid
decoupling was found to diminish and even disappear [10]. This is
another evidence that  there might be pressure (or temperature)
oscillations (high-frequency fluctuations of waves) during the
imposing process of high pressures. \newline Note that Kim and Chan
have also looked for this annealing effect by cooling a number of
solid samples from the liquid-solid coexistence region down to the
lowest temperature at a cooling rate that is up to 5 times slower
than that of Rittner and Reppy [10]. Kim and Chan, however, found
the supersolid fraction due to different annealing procedure can
differ by at most 15\%. But, they have not been able to eliminate
the superflow in any of the more than 50 bulk solid samples they
have studied so far in our laboratory. This fact could be explained
as the almost adiabatic process (smooth or long-period annealing) is
equivalent to a rather-low-frequency forcing (fluctuations of waves)
such that the small excitations being damped out and the frequency
range  is already beyond the triggering regime of phase transition
[14]. The latter (annealing effect) could also be traced in [15] or
partly related to that reported in [8].
{\it Acknowledgements}. The author is partially supported by the
2005-XJU-Scholars Starting Funds.

\end{document}